# Ultra-low-loss slow-light thin-film lithium-niobate optical modulator

Chenlei Li[1], Jianghao He[1], Ming Zhang[1], Yeyu Tong[2], Weixi Liu[1], Siyuan Wang[1], Lijia Song[1], Hongxuan Liu[1], Hengzhen Cao[1], LiuLiu[1,3], Yaocheng Shi[1,3], Daoxin Dai[1,3]*

1 State Key Laboratory for Extreme Photonics and Instrumentation, College of Optical Science and Engineering, International Research Center for Advanced Photonics, Zhejiang University, Zijingang Campus, Hangzhou 310058, China

[2]Microelectronics Thrust, The Hong Kong University of Science and Technology (Guangzhou), China.

[3]Jiaxing Key Laboratory of Photonic Sensing & Intelligent Imaging, Intelligent Optics & Photonics Research Center, Jiaxing Research Institute, Zhejiang University, Jiaxing 314000, China

*Corresponding Author: dxdai@zju.edu.cn

**Abstract**: Electro-optic modulators for next-generation optical interconnects require low loss-efficiency products, compact footprints, high modulation efficiency, broad bandwidths, and low losses. Here we propose and demonstrate a low-loss high-efficiency thin-film lithium-niobate Mach–Zehnder modulator enabled by a novel ultralow-loss slow-light structure based on apodized gratings in cascade. The present loss-engineered slow-light structure achieves excess losses as low as 0.6 dB/mm experimentally, which is tens of times lower than conventional slow-light structures, and a high modulation bandwidth up to 320GHz in theory is achieved with optimally-designed capacitively-loaded traveling-wave electrodes. Experimentally, the fabricated slow-light modulator with a 2.8-mm-long modulation region has an ultra-low loss-efficiency product $\alpha V_\pi L$ of 7.4 V·dB and a flat electro-optic response up to 67 GHz, enabling 100-Gbps on-off keying with high ERs of 4.5 dB at a low driving voltage of $2V_{pp}$, while 200-Gbps PAM4 and 150-Gbps PAM8 signals are also generated to show great promise for advanced modulation formats. In particular, it has also achieved the highest figure-of-merit (FOM) of ~186 Gbps·(dB/V)/(V·cm) for high-speed optical modulation, where FOM is defined as BR×(ER/$V_{pp}$)/($V_\pi L$) for comprehensively evaluating the modulator performances, including the bit rate (BR), the extinction ratio (ER) normalized with respective to $V_{pp}$, the modulation efficiency ($V_\pi L$). The outstanding performance of the present apodized-grating-based slow-light modulator shows great potential and paves the way for developing high-speed optical interconnects for both data-centers and high-performance computing systems.

**Keywords**: Lithium niobate; electro-optic; modulator; slow light; loss.

## Introduction

The relentless growth of data traffic is driving the need for optical interconnects with unprecedented capacity and efficiency[1-5]. To keep pace with this demand, optical interconnects are moving towards Terabit-per-second scale throughput on a single chip. Meeting these requirements calls for ideal electro-optic (EO) modulators with low loss-efficiency products ($\alpha V_\pi L$), compact footprints, high modulation efficiencies, low propagation losses, and broad bandwidths[2]. Such high-performance modulators are crucial for a variety of applications, including datacenters, co-packaged optics for switching, and computercom applications. Among various EO modulation approaches[6-15], Thin-film lithium-niobate (TFLN) has emerged as a promising platform potentially for high-performance EO modulators, offering tight optical confinement, low propagation loss, and a reliable Pockels effect[15-27]. However, the relatively low EO coefficient of lithium niobate ($\gamma_{33} \approx 27$ pm/V) usually limits the achievable modulation efficiency ($V_\pi L$) to 2.2-2.8 V·cm[15]. As a result, TFLN modulators typically require device lengths of 10-20 mm to achieve low drive voltages, leading to increased microwave attenuation and eventually compromising the electro-optic bandwidth. For example, increasing the device length by ~5 mm can decrease the bandwidth by ~30 GHz[15].

Various approaches have been explored to enhance the modulation efficiency and reduce the device footprint, including microring resonators[20], photonic crystals (PhC) cavities [21], Fabry-Pérot (FP) cavities [18, 22], slow-light waveguides [23, 28-32], etc. More details for the performances of these optical modulators are given in Table S1 of Supplementary Note I. Among them, slow-light modulators are particularly promising, as they can significantly enhance the interaction between the optical field and the modulating electric field by slowing the group velocity[33-37]. However, conventional slow-light waveguides based on e.g. photonic crystals, usually suffer from inherent drawbacks. They typically exhibit strong wavelength dependence, as the strong slow-light effect is restricted to the band edges. More importantly, the increased light-matter interaction in these structures often leads to excessive propagation losses, limiting the achievable bandwidth and efficiency for modulation. For example, in [31] when the enhancement factor for the group index of the slow-light structure increases from 2.4 to 5.5, the optical propagation loss significantly increases from 1.9 dB to more than 20 dB, and the E-O bandwidth greatly drops from 24.9 GHz to 10 GHz. Alternatively, slow-light modulators based on uniform gratings in cascade were proposed with flat-top spectral responses and simplified designs [23, 28, 38, 39]. Recent demonstrations of slow-light Mach-Zehnder modulators (MZMs) on silicon[28] and TFLN [23] have shown impressive performance with high bandwidths (>110 GHz or ~50 GHz) and low $V_\pi L$ values (0.96 V·cm or 1.29 V·cm). However, these devices still require high drive voltages (5 $V_{pp}$ on Silicon[28] or 8.5 $V_{pp}$ for TFLN [23]) even for achieving relatively low extinction ratios of 2 or 3.1 dB, as the length of the modulation region is not allowed to be extended to the millimeter-scale due to the very high substantial propagation loss of 5.4 dB/mm on Silicon or 13.3 dB/mm on TFLN. Therefore, it is still very challenging to make slow-light MZMs excelling in all the key properties such as low losses, compact footprints, low driving voltages, and particularly high bandwidths.

In this work, we propose and experimental demonstrate a novel slow-light modulator architecture offering high electro-optic bandwidth far exceeding 67 GHz with ultra-low propagation loss of 0.6 dB/mm, realized by introducing a unique design based on apodized gratings (instead of conventional uniform gratings). In this way, the grating is apodized gradually and thus the mode-mismatch loss is significantly reduced. Furthermore, the optical field in the grating is enhanced well in the regions with the shallowest grating corrugation, which greatly alleviates the light interaction with the grating corrugation during the light propagation, thus enabling significant reduction of the scattering loss. Besides, the high modulation bandwidth has been achieved by optimizing capacitively-loaded traveling-wave electrodes (CLTW), enabling effective velocity and impedance matching between the RF signal and the optical signal. For the fabricated slow-light MZM designed with an extended device length of 2.8 mm, a loss-efficiency product $\alpha V_\pi L$ as low as 7.4 dB·V is achieved as the best one for slow-light modulators, to the best of our knowledge. Finally, we successfully demonstrate high-speed data transmissions with the data rate of 100 Gbps (OOK), 200 Gbps (PAM-4), and 150 Gbps (PAM-8), showing decent dynamic extinction ratios of 4.5, 3.8, and 3.6 dB even with a peak-to-peak voltage $V_{pp}$ of only 2.0 V. In order to comprehensively evaluate optical modulators which are always desired to have high bit rates (BRs), high extinction ratios (ERs) normalized with respective to $V_{pp}$ and low $V_\pi L$ (high modulation efficiency), here we define a special figure-of-merit (FOM) as BR×(ER/$V_{pp}$)/($V_\pi L$), and the present slow-light modulator has an FOM of ~186 Gbps·(dB/V)/(V·cm), which is the highest among all reported optical modulators. The outstanding performance of the present apodized-grating-based slow-light MZM shows great potential and paves the way for developing high-speed optical interconnects available for both data-centers and high-performance computing systems.

**Results**

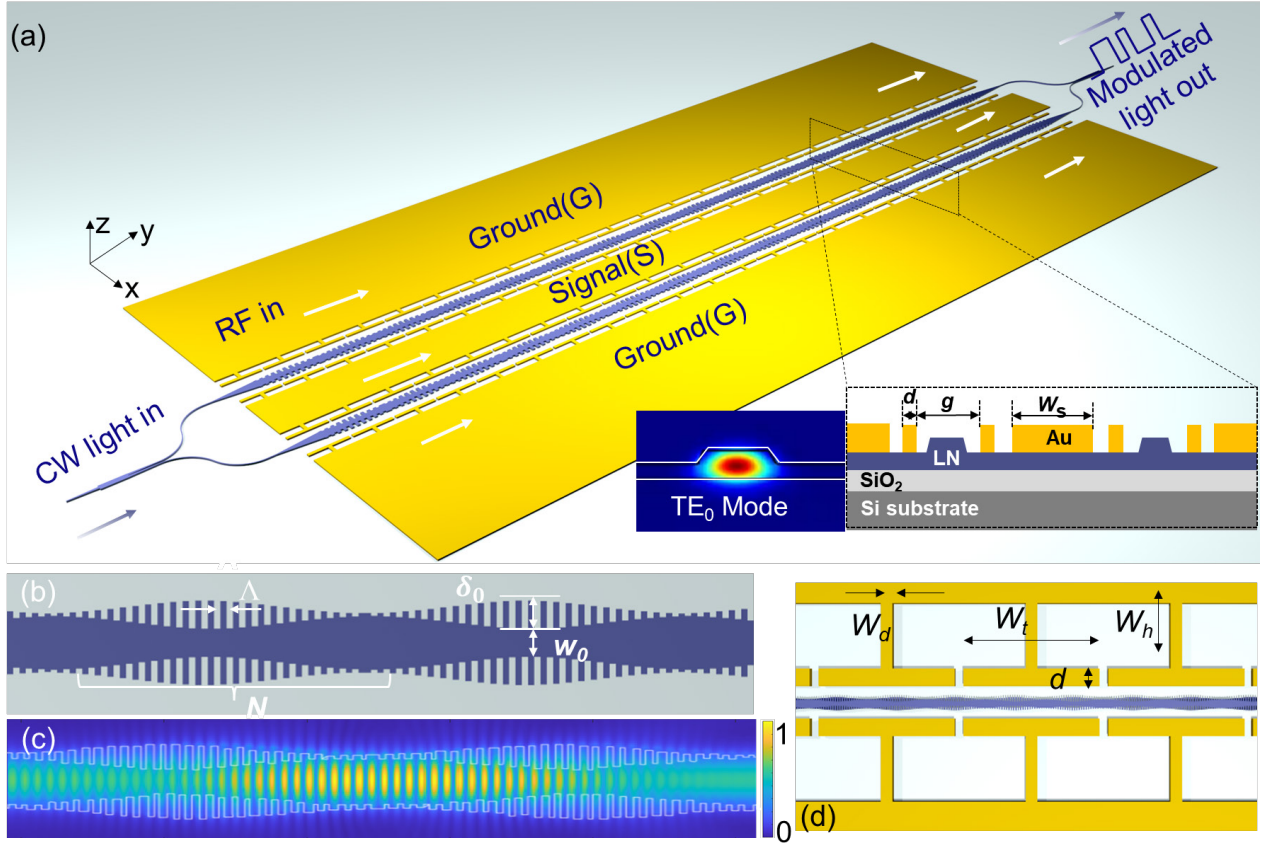

Fig. 1 (a), Schematic configuration of the proposed slow-light modulator. inset: left, simulated mode profile in the designed LN photonic waveguide; right, cross-sectional schematic of the nanophotonic LN modulator. (b), Top view of the coupled Bragg resonator based slowlight waveguide that consists of series of Bragg-gratings separated by a π-phase shifter region. (c), Simulated light propagation in the designed slow-light waveguide. (d), Schematic view of the proposed periodic capacitively loaded traveling-wave (CLTW) electrode structure.

According to the modulation efficiency $V_\pi L$ of a TFLN slow-light MZM with a push-pull configuration is given as:

$$V_\pi L = \frac{\lambda \cdot g}{2 \cdot \Gamma \cdot n_g \cdot n^2 \cdot \gamma_{33}}, \tag{1}$$

where λ is the wavelength, *n* is the refractive index, $\gamma_{33}$ is the EO coefficient, and *g* is the electrode gap, $\Gamma$ represents the EO overlap factor of the optical mode, $n_g$ is the group index. Considering the limitation posed by reducing the electrodes spacing on the modulation-efficiency improvement, enhancing the group index is an effective approach to improve the modulation efficiency. However, slow-light waveguides based on conventional uniform gratings suffer from significant scattering losses as well as mode-mismatch losses, limiting the allowable interaction length and consequently the achievable dynamic extinction ratio (ER) [23, 28]. The high loss in conventional slow-light waveguide is attributed to two main factors [40-45]: One is the enhanced optical scattering loss due to light interaction with the grating corrugation during the light propagation, especially near the π-phase-shift regions. The other one is the mode mismatch loss between the grating and non-grating sections. Addressing these loss mechanisms is essential for realizing efficient, low-drive-voltage slow-light modulators with extended interaction lengths and improved performance.

To overcome these limitations, we leverages an ultralow-loss slow-light waveguide architecture to achieve a electro-optic (EO) modulator with low loss-efficiency products ($\alpha V_\pi L$), compact footprints, high modulation efficiencies, low propagation losses, and broad bandwidths as shown in Fig. 1(a-d). For the present slow-light waveguide, two adjacent apodized gratings are separated by a π-phase shifter to form a resonator, as shown in Fig. 1(b). The number of cascaded resonators is denoted by $N_p$. The apodized-grating corrugation depth $\delta$ is defined by the following Gaussian function [46, 47] to achieve a gradual change in the mode field within the slow light waveguide,

$$\delta = \delta_0 \exp\left[-b\left(\frac{i}{N} - \frac{1}{2}\right)^2\right], \qquad (2)$$

where $\delta_0$ is the offset maximum, $b$ is the apodization strength, and $N$ is the number of grating teeth. The width for the central part of the waveguide is $w_0$. Our simulation reveals that the optical field is strongly localized and enhanced in the low-loss π-phase-shift regions (Fig. 1(c)) as discussed before[48].

This gradual grating apodization significantly reduces the mode-mismatch losses during propagation, particularly as the optical field enhancement in the π-phase-shift regions—where the grating corrugation is shallowest—substantially mitigates light interaction with the grating corrugations, and thus, decreases scattering losses caused by sidewall roughness due to fabrication errors (see more details in Supplementary Note I). Additionally, the present design with apodized gratings also elegantly minimizes the mode mismatch between the slow-light waveguides and the regular waveguides, while the design with conventional uniform gratings has abrupt transitions leading to additional mode mismatch losses. It should be noted that the mode gradient resulting from the intrinsic apodization in the slow-light waveguide differs from the approach described in ref. [29]. While an additional linear taper was introduced to reduce mode mismatch, it inadvertently acted as an additional reflector at both ends of the slow light waveguide, thereby inducing excess loss. Consequently, our design strategy enables a significant reduction in the scattering losses, facilitating efficient electro-optic modulation without excessive propagation losses.

The proposed slow-light waveguides are designed with a 400 nm-thick X-cut LN thin film in this work, an air upper-cladding, and a 3-μm-thick $SiO_2$ under-cladding layer, while the ridge height is designed to be 200 nm. The mode profile of the TE fundamental mode in the designed LN photonic waveguide is shown by inset in Fig. 1(a) (left). The designed slow-light MZM works with the push–pull configuration, and the cross-section of the modulation region is given by the inset in Fig. 1(a) (right).

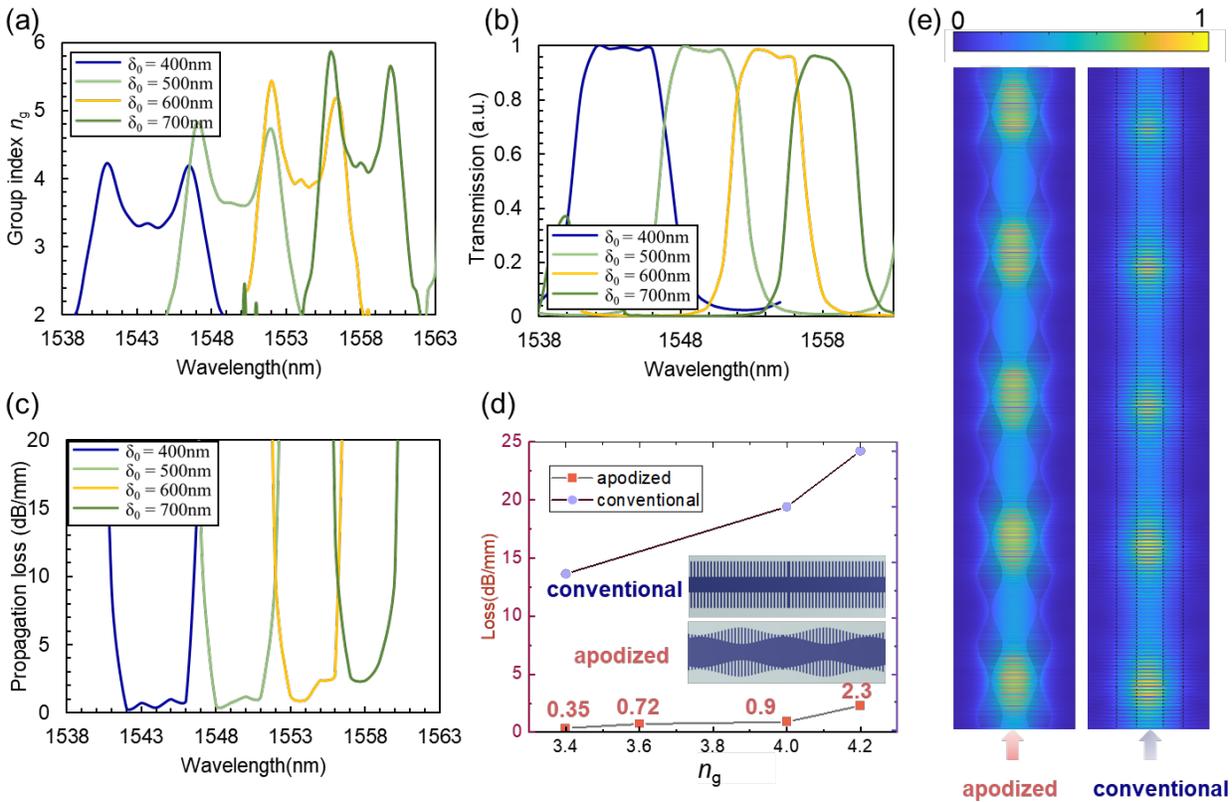

Fig. 2 Calculated results of the apodized-grating slow-light waveguide: (a), group index $n_g$. (b), transmission spectral responses. (c), calculated propagation loss. (d), calculated losses for the apodized-grating and conventional slow-light waveguide. (e), simulated electric field intensity distributions inside the apodized-grating and conventional slow-light

waveguide waveguides with $n_g \approx 4.5$.

To further understand the proposed slow-light waveguide's effect on improved performance, we investigate the key characteristics of the structure. Fig. 2(a-c) presents the calculated group index $n_g$, transmission spectral and the propagation loss for the proposed slow-light waveguide with varying grating corrugation depths $\delta_0$. The grating parameters are given as follow: the grating period $\Lambda$= 438 nm, the apodization strength $b$ = 10, the grating teeth number $N$= 30, the number of resonators in cascade $N_p$=3, the small $N_p$ was chosen to avoid computational complexity. Here the $\delta_0$ is varied from 400 to 700nm, respectively, while the center width $w_0$ is fixed as 400 nm. The apodized grating design achieves remarkably low propagation losses of 0.35, 0.72, 0.9, and 2.3 dB/mm for $n_g$ of 3.4, 3.6, 4.0, and 4.3, respectively, indicating that the grating corrugation in the field-enhancement regions does not introduce significant excess losses (ELs). Fig. 2(d) give a comparison about the ELs between the proposed slow-light waveguide based on apodized gratings and the conventional structure (corresponding $b$=0) used in [23] (more details are shown in Supplementary Note II). The conventional slow-light waveguide exhibits pretty high propagation losses up to 14, 20, and 25 dB/mm for group indices of 3.4, 3.8, and 4.1, respectively. In contrast, the excess loss of the present design with apodized gratings is reduced greatly by more than an order of magnitude compared to the conventional one. Fig. 2(e) shows the simulated light propagation in the present (left) and conventional (right) slow-light waveguides designed with $n_g$ =~4.5, respectively. The apodized-grating design maintains low propagation losses even with enhanced $n_g$, while the conventional one suffers from significant optical attenuation.

To balance the optical bandwidth and the modulation efficiency, the structural parameters are chosen as $w_0$ = 0.6μm, $\delta_0$ = 0.6μm, $N_p$= 80, and $N$ = 40. This configuration yields a group index of $n_g$~3.8, an 3dB optical bandwidth of ~5.5 nm, and minimal theoretical ELs of 0.3 dB/mm. The dependence of ELs, $n_g$, and optical bandwidth on the parameters ($N_p$, $N$, $\delta_0$, and $w_0$) is further analyzed in Supplementary Note III. The ultralow-loss performance enables a 2.8-mm-long modulation region, which is impractical for conventional slow-light waveguides due to their prohibitively high ELs. The MZM incorporates a ground-signal-ground (GSG) traveling-wave electrode configuration with CLTW (Fig. 1(d)) to achieve velocity matching between the electrical and optical signals. T-shaped periodic structures are introduced between the ground and signal electrodes to minimize the electrode gap, maintain impedance matching, and reduce the electrical signal velocity (see Supplementary Note IV for details). Simulations indicate an electro-optic (EO) bandwidth exceeding 320 GHz for the 2.8-mm-long device. The ultralow-loss slow-light waveguide design enables extended interaction lengths, allowing for high dynamic extinction ratios (ERs) even at relatively low drive voltages

Fig. 3(a-d) shows the fabricated TFLN slow-light modulators (see more details about the fabrication in Supplementary Note V). The apodized grating design, with reduced corrugation strength towards the resonator edges, eliminates the need for transition regions between the straight and slow-light waveguides (Fig. 3(d)). The half-wave voltage measurements the fabricated Mach-Zehnder modulator (MZM) is characterized by applying a 50-kHz triangular wave signal from an arbitrary function generator (AFG) and measuring the output with an oscilloscope (DSO) via a photodetector (PD) (Fig. 3(e)). Sweeping the DC voltage from -4 V to 4 V reveals a half-wave voltage ($V_\pi$) of 4.4 V, corresponding to a modulation efficiency ($V_\pi L$) of 1.23 V·cm, in good agreement with simulations.

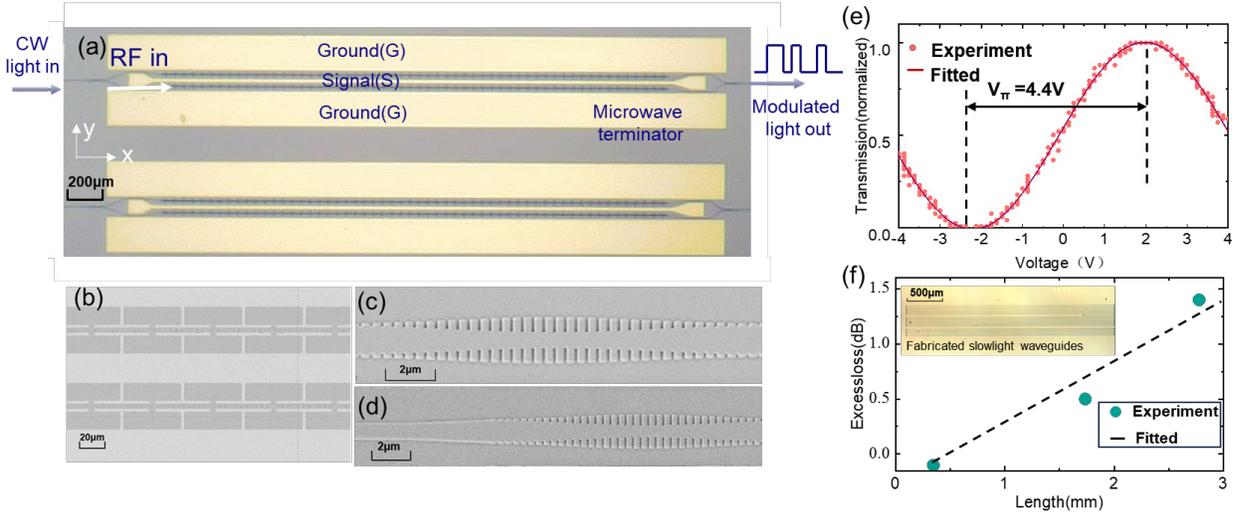

Fig. 3 (a), Optical microscope images of fabricated slow-light MZMs. Scanning electron microscopy pictures of the modulation section (b), the apodized Bragg-grating (c) and the waveguide junction between the regular waveguide and the slow-light waveguide (d). (e), Normalized optical transmissions of the fabricated LNOI photonic chip as a function of the applied voltage. f, Static extinction ratio (ER) as a function of the applied voltage. (f) Measured propagation losses for waveguides with $N_p$ = 10, 50, and 80.

The optical properties of the slow-light waveguides are further investigated (Supplementary Note VI). The measured 3-dB optical bandwidth is ~6 nm from 1550.3 to 1556.6 and remains insensitive to the number of cascaded resonators $N_p$. As the number of grating teeth $N$ increases, the propagation loss remains nearly constant, while the slow-light effect is enhanced, and the optical bandwidth gradually reduces from 6 nm to 2 nm, consistent with previous reports [35]. Fig. 3(f) shows the measured propagation losses for waveguides with $N_p$ = 10, 50, and 80, and $N$ = 40, corresponding to a $n_g$ of 3.8. The inset shows an image of the fabricated slow-light waveguides. Remarkably, the present design achieves a low propagation loss of only ~0.6 dB/mm, which is more than an order of magnitude lower than conventional slow-light waveguides [23]. The results were normalized with the straight waveguide fabricated on the same chip near the slow-light waveguides. Notably, the waveguide structure was patterned using electron-beam lithography (EBL), which often introduces stitching errors due to the limited writing field and thus additional ELs. Therefore, further performance improvements can be expected by employing standard ultraviolet stepper lithography and optimizing the etching process to reduce sidewall roughness and scattering losses. The fabricated MZM achieves an ultra-low $\alpha V_\pi L$ of 7.4 dB·V, making it the best-performing slow-light modulator reported to date.

Fig. 4(a) shows the measured EO responses $S_{21}$ of the fabricated MZM when operating at different operation wavelengths of 1550.5, 1553.5, and 1556.1 nm, respectively, using the experimental setup described in Supplementary Note VII. The frequency range is limited to 67 GHz due to the Lightwave component analyzer (LCA) constraints. Notably, the EO responses remain flat without roll-off within the measured frequency range for all operating wavelengths spanning the 6-nm 3-dB optical bandwidth, in excellent agreement with theoretical predictions. The slow-light MZM is further evaluated for high-speed data transmission. Fig. 4(b-d) show the measured eye diagrams for 100 Gbps on-off keying (OOK), 200 Gbps four-level pulse amplitude modulation (PAM4), and 150 Gbps eight-level pulse amplitude modulation (PAM8) signals, all driven with a peak-to-peak voltage ($V_{pp}$) of 2.0 V. The eye diagrams exhibit wide openings and high extinction ratios of 4.5, 3.8, and 3.6 dB for the 100 Gbps OOK, 200 Gbps PAM4, and 150 Gbps PAM8 signals, respectively, confirming the modulator's excellent performance at high data rates. The energy efficiency of the slow-light MZM is estimated to be 40 fJ/bit for 200 Gbps PAM4 signaling. Remarkably, the power consumption can be further reduced by extending the modulation region length, leveraging the ultralow excess losses (ELs) of the slow-light waveguide design.

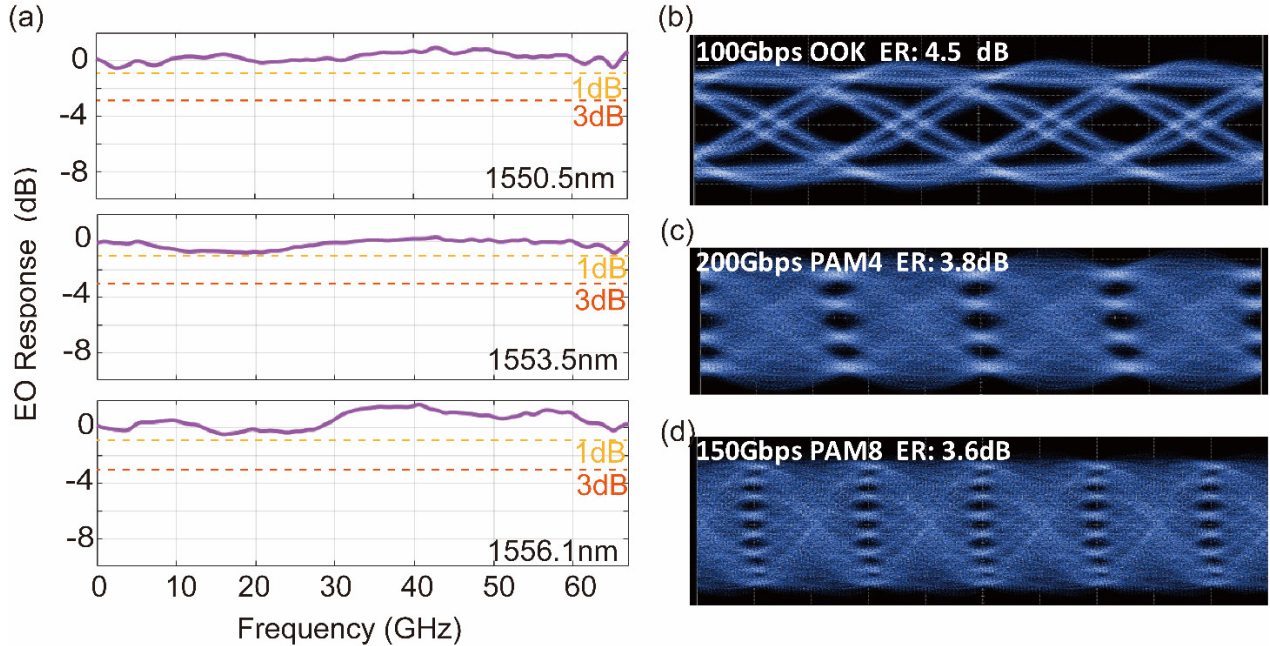

Fig. 4 (a) Measured EO responses for the fabricated slow-light modulator under different working wavelengths spanning from 1550.5nm to 1556.1nm. (b-d), Measured optical eye-diagrams at data rates of 100 Gbps OOK, 200 Gbps PAM4, 150 Gbps PAM8.

**Discussion**

In this work, we have proposed a novel approach to overcome the long-standing challenge of achieving low-loss, compact, and high-performance EO modulators by introducing novel apodized gratings in cascade for slow-light waveguides. The key advantage of the present design lies in its ultra-low propagation loss of 0.6 dB/mm, which is over an order of magnitude lower than conventional slow-light waveguides. The present slow-light waveguide with apodized gratings is allowed to directly connect to the input/output straight waveguides without any additional mode converters. In particular, the slow-light strategy enabling ultra-low losses allows to use a relatively long modulation region (which is not achievable for conventional slow-light modulators with high losses), which consequently helps to achieve improved dynamic ERs and lowered drive voltages, paving the way for high-performance EO modulators even supporting advanced modulation formats such as PAM-4 and PAM-8. The present slow-light MZM has a high modulation efficiency of 1.2 V·cm, which is about twice higher than regular MZMs without slow-light structures. In addition, the periodic CLTW electrodes have been utilized to overcome the mismatch of the impedance and velocity for the electrical and optical signals, greatly enhancing the modulation bandwidth to be far beyond 67 GHz (the maximal bandwidth of the equipment).

Compared with those EO modulators based on e.g. microring resonators or photonic-crystal cavities, the present slow-light MZM shows a broad optical bandwidth of 6 nm, greatly relaxing the critical requirement for the operation-wavelength alignment and making the device robust to external environmental variations, thus reducing the power consumption for the resonance-wavelength control. Such compact and broadband phase-modulation also provides a promising path to be scaled for wavelength-division multiplexing (WDM) and coherent modulation techniques, enabling ultra-high link capacity with multiple optical modulators integrated on a single chip.

Table 1 gives a summary for the reported devices, including the loss-efficiency product $\alpha V_\pi L$, the data BR, the normalized ER and the EO bandwidth. It can be seen clearly that silicon EO modulators are still necessary to be improved regarding to the high losses as well as the limited EO bandwidth, which thus hinders to be used for satisfying the ever-increasing demand for ultra-high-BR data transmitting. In contrast, TFLN EO modulators indeed provide a promising option with reduced losses and improved bandwidths, while the modulation efficiency is still low due to the limited EO coefficient of LN and thus centimeter-scale modulation regions are usually necessary for lowering the voltage $V_{pp}$. As demonstrated previously, with the assistance of slow-light structures, one can enhance the modulation efficiency and shrink the footprint for the EO modulators, which however still suffer the issues related with high propagation losses, limited EO bandwidths as well as narrow optical bandwidths. As shown in Table 1, our fabricated TFLN slow-light modulator with optimized parameters have highlighted an ultra-low value of $\alpha V_\pi L$ (~7.4 dB·V) among the reported high-efficiency modulators, with low losses of 0.6 dB/mm, a high modulation efficiency $V_\pi L$ of 1.2V·cm, a compact footprint of ~2.8 mm, low driving voltage (i.e., low power consumption) of 2 V, as well as high optical bandwidths of ~6 mm simultaneously. Experimentally, we have obtained a flat EO response up to 67GHz while the calculated 3-dB EO bandwidth reaches 320 GHz, showing the great potential for ultra-high-speed data modulation. Here data transmissions with 100 Gbps OOK, 200 Gbps PAM4, and 150 Gbps PAM8 signals with high ERs (4.5 dB,3.8 dB,3.6dB) have been demonstrated successfully, showcasing the great potential to work with advanced modulation formats. Higher bit-rate beyond 200 Gbps is possible if advanced experimental setup is available.

In order to give a comprehensive evaluation for the over-all performance of high-speed optical modulators, here we define a special figure-of-merit (FOM) as the product of the bit rate (BR), the normalized extinction ratio (ER) and $1/(V_\pi L)$, i.e., FOM=BR×(ER/$V_{pp}$)/($V_\pi L$). A higher FOM is obtained for an optical modulator with a higher BR, a higher normalized ER as well as a lower $V_\pi L$ (higher modulation efficiency). Fig. 5 shows the calculated FOM for all the optical modulators reported. As it can be seen, the FOM of the most reported optical modulator is no more than 100. One of the representative slow-light silicon modulators is the one reported with 110 GHz in [28], which has a loss as high as 29.8 dB/mm and an FOM of 65 Gbps·(dB/V)/(V·cm). Another example is the slow-light TFLN modulator reported in [23], showing a loss of 13.3 dB/mm and an FOM of 14 Gbps·(dB/V)/(V·cm). In contrast, the present slow-light modulator works with a low loss of 0.6 dB/mm and an FOM as high as 182 Gbps·(dB/V)/(V·cm), which is a new record. From the comparison given in Fig. 5 and Table 1, it can be seen that the present slow-light modulator is an example with the best over-all performance to satisfy the demands in the real applications.

Further reducing the device loss is possible when using standard UV lithography (instead of E-beam lithography used here) because the stitching errors and the sidewall roughness can be minimized. Improving the dynamic ER for higher-order modulation formats can be addressed through electronic and photonic equalization techniques [49-51]. Finally, co-integration with laser sources and coherent receiver circuitry will be very attractive for the deployment of next-generation optical communication systems.

As a summary, this groundbreaking slow-light modulator has been developed to overcome the trade-offs between the optical loss, the device size, the modulation efficiency, and the bandwidth that have hindered conventional EO modulators. By offering a new route to high-density, energy-efficient, and ultra-high-speed optical interconnects, our approach opens the door to Terabit/s-scale datacom and computercom applications. With further advances in modulation formats, this work paves the way for a new-generation EO modulators that drive the future of high-capacity optical communications and high-performance computing systems.

TABLE 1 Comparison of reported silicon and TFLN EO Modulators.

| Ref | Platform | Structure | α (dB/mm) | Length (μm) | $V_\pi L$ (V·cm) or Efficiency (pm/V) | $\alpha V_\pi L$ (dB·V) | Data rate (Gbps) | Normalized ER-OOK ER/$V_{pp}$ (dB/V) | FOM BR×(ER/$V_{pp}$)/$V_\pi L$ | EO BW (Hz) |
|---|---|---|---|---|---|---|---|---|---|---|
| [50] | SOI | Regular MZM | 2.2 | 2000 | 1.4 | 30.8 | 90-OOK 100-PAM4 | 0.66 @ 90 Gbps | 42 | 58 G |
| [51] | SOI | Regular MZM | 2.4 | 1500 | 1.6 | 38.4 | 200-PAM4 | / | / | 22.5 G |
| [15] | TFLN | Regular MZM | 0.03 | 5000 | 2.2 | 0.66 | / | / | / | 100 G |
| [15] | TFLN | Regular MZM | 0.03 | 20000 | 2.3 | 0.69 | 210,8-ASK; 70-OOK | / | / | 45 G |
| [25] | TFLN | Regular MZM | 0.1 | 3000 | 2.2 | 2.2 | 100-OOK; 112-PAM4 | 1.25 @100 Gbps | 56 | 70 G |
| [27] | TFLN | Regular MZM | 0.24 | 12500 | 2.1 | 5.0 | 168-PAM8, 56-OOK | 4.0 @ 56Gbps[a] | 107 | 70 G |
| [20] | TFLN | Regular MZM | 0.3 | 2000 | 1.8 | 5.4 | 22 | 1.42 @ 22Gbps | 5.8 | 15 G |
| [30] | SOI | Slow light | 12.3 | 162 | 0.18 | 22.1 | 20-OOK | 0.71 @ 20Gbps | 77 | 28 G |
| [28] | SOI | Slow light | 29.8 | 124 | 0.96 | 286.1 | 112-OOK | 0.63 @100 Gbps | 65 | 110 G |
| [23] | TFLN | Slow light | 13.3 | 370 | 1.29 | 171.5 | 80-OOK | 0.23@ 80Gbps | 14 | 50 G |
| [20] | TFLN | MRR | 0.3 | 300[b] | 7 pm/V | / | 40-OOK | / | / | 30 G |
| [31][c] | TFLN | Slow light | 12 | 820 | 0.67 | / | 60-OOK | 0.53 @ 60Gbps | 47 | 10 G |
| [21] | TFLN | PhC cavity | / | 30 | 16 pm/V | / | 11-OOK | / | / | 17.5 G |
| [18] | TFLN | FP cavity | / | 50 | 7 pm/V | / | 100-OOK 140-PAM4 | 0.7 @100 Gbps | / | 110 G |
| This work | TFLN | Slow light | 0.6 | 2800 | 1.23 | **7.4** | 200-PAM4 100-OOK | 2.25 @100 Gbps | **182** | >> 67G (expr.) 320G (theory) |

[a] data using offline digital signal processing algorithms was not considered here
[b] estimated from the figure
[c] parameter shown in the table corresponding to a wavelength @1548.3nm

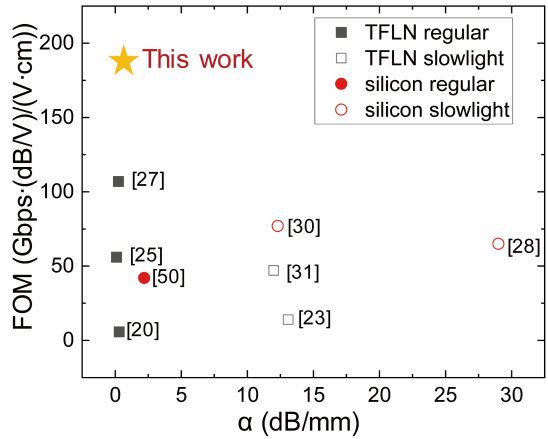

Fig. 5 The FOM for TFLN and silicon MZMs reported. Here FOM=BR×(ER/$V_{pp}$)/($V_\pi L$)

**Research funding**: This work was funded in part by National Natural Science Foundation of China (NSFC) (62405271, 62205286, 92150302, U23B2047, 62321166651, 62135010), China Postdoctoral Science Foundation (2023M733039), Natural Science Foundation of Zhejiang Province (LQ21F050006), the Innovation Program for Quantum Science and Technology (2021ZD0301500).
**Disclosures.** The authors declare no conflicts of interest.